\documentclass[reprint,superscriptaddress,preprintnumbers,amsmath,amssymb,floatfix]{revtex4-1}
\usepackage{graphicx}
\usepackage{amsmath}
\usepackage{gensymb}
\usepackage{mathtools}
\usepackage{comment}

\begin{document}
\title{Multimode internal resonances in a MEMS self-sustained oscillator}
\author{S. Houri} \email{samer.houri.dg@hco.ntt.co.jp}
\affiliation{NTT Basic Research Laboratories, NTT Corporation, 3-1 Morinosato-Wakamiya, Atsugi-shi, Kanagawa 243-0198, Japan}
\author{D. Hatanaka}
\affiliation{NTT Basic Research Laboratories, NTT Corporation, 3-1 Morinosato-Wakamiya, Atsugi-shi, Kanagawa 243-0198, Japan}
\author{M. Asano}
\affiliation{NTT Basic Research Laboratories, NTT Corporation, 3-1 Morinosato-Wakamiya, Atsugi-shi, Kanagawa 243-0198, Japan}
\author{H. Yamaguchi}\email{ hiroshi.yamaguchi.zc@hco.ntt.co.jp}
\affiliation{NTT Basic Research Laboratories, NTT Corporation, 3-1 Morinosato-Wakamiya, Atsugi-shi, Kanagawa 243-0198, Japan}

\date{\today}

\begin{abstract}
We investigate the dynamics of a microelectromechanical (MEMS) self-sustained oscillator supporting multiple resonating and interacting modes. In particular, the interaction of the first four flexural modes along with the first torsional mode are studied, whereby 1:2, 1:3, and 2:1 internal resonances occur. Even and odd modes are induced to couple by breaking the longitudinal symmetry of the structure. Self-oscillations are induced in the second flexural mode via a gain-feedback loop, thereafter its frequency is pulled into a commensurate frequency ratio with the other modes, enabling the oscillator to act as a driver/pump for four modes simultaneously. Thus, by leveraging multiple internal resonances, a five modes frequency-locked comb is generated.
\end{abstract}

\maketitle

\indent\indent\ Micro- and nano-electromechanical devices (MEMS/NEMS) offer an excellent platform for the experimental investigation of nonlinear dynamical phenomena \cite{lifshitz2008nonlinear} as they offer systems with multiple orders of nonlinearities and multiple interacting modes \cite{houri2017direct,antonio2012frequency,arroyo2016duffing,mangussi2016internal,guttinger2017energy,shoshani2017anomalous,taheri2017mutual,zanette2018stability,czaplewski2018bifurcation,czaplewski2019bifurcation,houri2019limit,westra2010nonlinear,ganesan2017phononic,ganesan2018phononic,cao2014phononic}. In particular, the investigation of internal resonance, whereby nonlinearities couple two or more vibrational modes of a device, has seen significant recent interest \cite{antonio2012frequency,arroyo2016duffing,mangussi2016internal,guttinger2017energy,shoshani2017anomalous,taheri2017mutual,zanette2018stability,czaplewski2018bifurcation,czaplewski2019bifurcation,houri2019limit} with promising applications in the areas of mechanical frequency combs \cite{czaplewski2018bifurcation,ganesan2017phononic,ganesan2018phononic,cao2014phononic} and improved MEMS oscillators \cite{antonio2012frequency,taheri2017mutual,zanette2018stability}.\\
\indent\indent\ The inter-modal energy transfer that occurs as a result of this internal resonance was shown to impact dissipation in nanomechanical systems \cite{guttinger2017energy}, has been associated with the creation of limit cycles via a Hopf \cite{houri2019limit} or a SNIPER \cite{czaplewski2018bifurcation} bifurcations, and results in a reduced frequency noise in a MEMS oscillator \cite{antonio2012frequency}. The impact of internal resonance on self-sustained oscillators is a particularly interesting topic due to the potential gain in performance, where in addition to lower frequency noise, coherent energy transfer between the modes was observed \cite{chen2017direct}, and in a phase locked-loop (PLL) implementation, a large tuning range was achieved \cite{arroyo2016duffing}, and inter-modal synchronization demonstrated \cite{taheri2017mutual}. Despite these interesting developments, self-oscillating internally resonant systems remain vastly under explored compared to the case of driven systems. This is particularly true for systems that utilise gain feedback loops rather than PLLs. Questions regarding stability, tunability, and bifurcations  remain to be investigated, especially when considering an analytical treatment in which the coupling is fully nonlinear. Furthermore, the possibility of achieving multiple internal resonances simultaneously is unreported, despite its fundamental interest and its potential to form modal frequency combs.\\
\indent\indent\ This publication builds on previous work \cite{houri2019limit} where a two-mode cavity exhibiting internal resonance was investigated in an open loop configuration, i.e. driven by an external force, and showed regions of instabilities in the form of Hopf supercritical bifurcations. In contrast, this work aims to address some of the unexplored physics of internally resonant self-sustained oscillators implemented through a gain feedback loop. In particular, questions of tuning range are addressed experimentally as well as analytically, and the potential to achieve multiple internal resonances simultaneously is demonstrated by forcing both even and odd modes to interact, thus practically doubling the accessible parameter space.\\
\indent\indent\ The MEMS device used in this work is a  $\mathrm{AlGaAs/GaAs}$ heterostructure piezoelectric clamped-clamped beam. The structure, which is described in more detail in \cite{yamaguchi2017gaas,houri2019pulse}, is 150  $\mu$m long and possesses two electrically isolated actuation electrodes on each side of the  beam\textsc{\char13}s length near the anchoring points. A -1~V\textsubscript{DC} is applied to one of the electrodes while the other one is shorted, the dc is necessary to maintain a linear piezoelectric transduction and avoid the nonlinearities of the metal-semiconductor Schottky junction upon the application of an AC signal (see supplementary material). This single sided actuation makes it possible to efficiently transduce both even and odd modes. Furthermore, due to the DC voltage a constant strain is applied on one end of the beam, thus breaking the longitudinal symmetry and making it possible to couple even and odd modes.\\
\indent\indent\ The MEMS device is placed in a vacuum chamber ($\sim$ $10^{-4}$ Pa), actuated electrically and its motion measured using a laser Doppler vibrometer (LDV). The structure\textsc{\char13}s spectral response is obtained using a Zurich Instruments lock-in amplifier (HF2LI) and exhibits four resonance peaks between 0 and 1.6 MHz corresponding to the first four flexural modes. The resonance modes show a linear response when subjected to a low amplitude drive ($\sim$ 20 mV\textsubscript{PP}), with frequencies of 319, 519, 953, and 1564 kHz for the first through fourth modes respectively. Note that the frequency ratio between the 1st and 3rd and the 2nd and 4th modes are very close to 1:3, thus internal resonance for these modes can be expected for large drive amplitudes. In addition, the Brownian response of a torsional mode, that is not efficiently excited electrically, is observed around 1083 kHz, corresponding to a frequency ratio of 1:2 with the second flexural mode.\\
\indent\indent\ As the drive voltage is increased to moderate levels, all of the four flexural modes show a nonlinear Duffing response, of the softening type for the first mode and the hardening type for the other 3 modes as shown in Fig.~1(a). The linear and nonlinear (Duffing) parameters are fitted, also shown in Fig.~1(a). The fits were done following the procedure described in \cite{davidovikj2017nonlinear,houri2019limit}, the modal constants are summarized in supplementary material.\\
\indent\indent By further increasing the drive voltage, deviations from a Duffing frequency response are observed in the sweeps for the first and second modes indicating the onset of internal resonance. The fingerprint of internal resonance can be clearly identified by plotting the jump-down frequency of the Duffing as a function of the modal amplitudes as shown in Fig.~1(b) and Fig.~1(c) for the first and second modes respectively. The plots first follow a quadratic relation \cite{anderson2006nanomechanics}, but as the drive is increased the modal response deviates from this quadratic relation and forms plateau features indicating energy transfer between the lower and higher modes. For the first flexural mode only a 1:3 internal resonance is observed (with the third flexural mode), whereas for the second flexural mode both 1:3 (with the fourth flexural mode) and 1:2 (with the torsional mode) internal resonances are observed.\\
\begin{figure}
	\graphicspath{{Figures/}}
	\includegraphics[width=85mm]{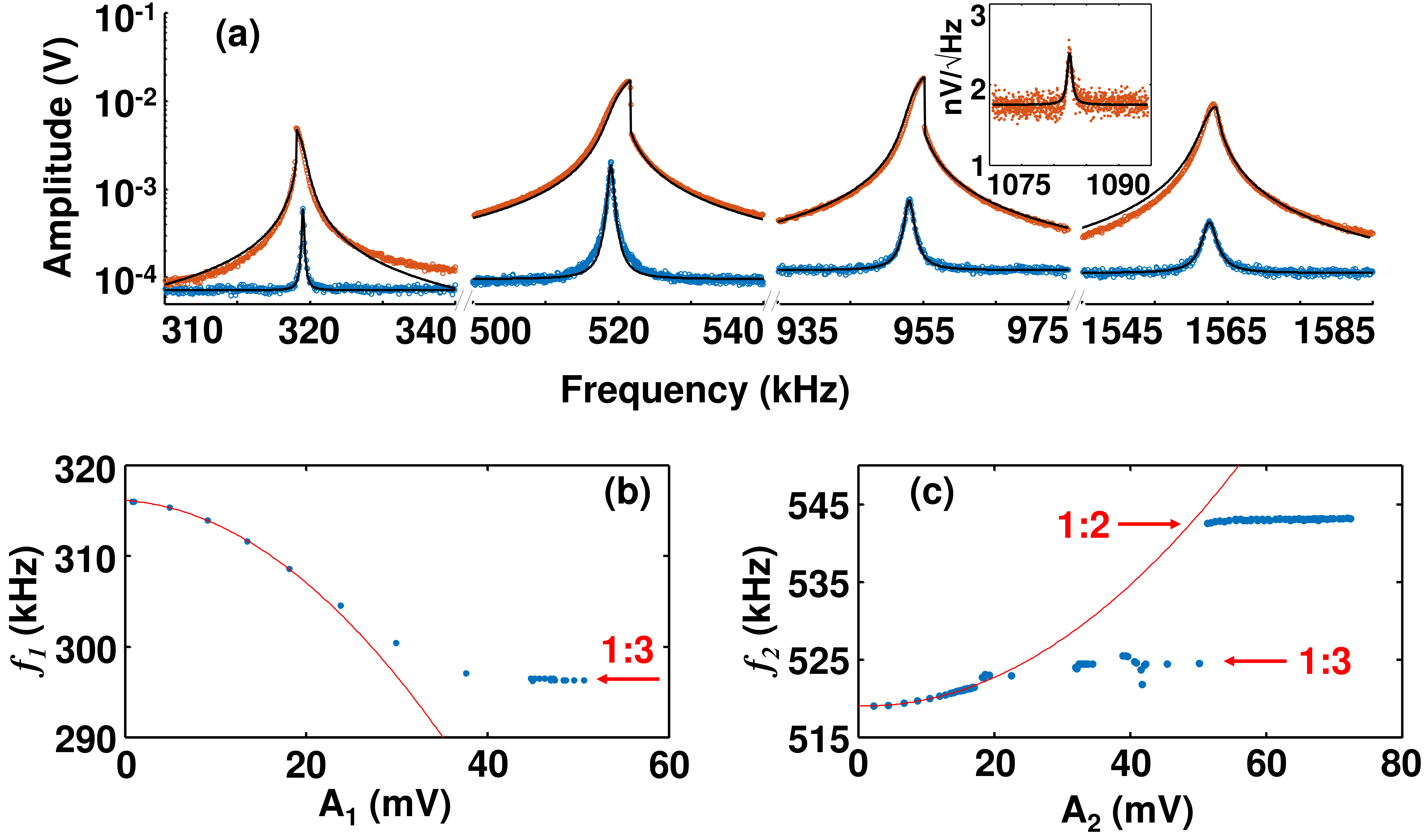}
	\caption{(a) Spectral response of the MEMS multimode cavity as obtained using a lock-in amplifier shown for low 10 mV (blue) and moderate 220 mV (red) excitation voltages, except for the fourth mode 40 mV and 1 V, respectively. Also shown is the PSD of the Brownian spectrum of the torsional mode (inset). The frequency-amplitude locus of the maxima of the Duffing response plotted for the first (b) and second (c) modes, respectively. The response first follows a quadratic relation
(red traces) then deviates forming plateaus due to internal resonance. The second mode shows two consecutive plateaus indicative of a 1:3 and a 1:2 internal resonance respectively.}
\end{figure}
\indent\indent\ To model the dynamics of these internal resonances, nonlinear interaction terms are introduced into a Duffing equation \cite{houri2019limit,czaplewski2019bifurcation}. Here the interaction terms Hamiltonian and the resulting coupled equations read as:\\
\begin{eqnarray}
\begin{rcases}
\mathrm{\mathcal{H}_{jk} =  \kappa_{jk}x_k^2x_j + g_{jk}x_k^3x_j }\\
\mathrm{\ddot{x}_k + (\gamma_k + \beta_kx_k^2)\dot{x}_k + \omega_{k,0}^2x_k + \alpha_kx_k^3 +\frac{\partial \mathcal{H}_{jk}}{\partial {x}_k} = F_k}\\
\end{rcases}
\end{eqnarray}
\indent\indent where ${\mathcal{H}_{jk}}$ is the modal interaction Hamiltonian between modes \textit{j} and \textit{k}, $\kappa_{jk}$ and $g_{jk}$ are the second and third order nonlinear coupling parameters (corresponding to three-wave and four-wave mixing respectively), $x_k$ and $x_j$ are the modal displacements, $\gamma_k$ and $\beta_k$ are the modal linear and nonlinear (van der Pol) damping parameters, $\omega_{k,0}$ and $\alpha_k$ are the natural frequency and Duffing nonlinearity, respectively, and $F_k$ is the modal forcing term. Dispersive mode coupling \cite{westra2010nonlinear} is dropped since its value is small compared to other terms \cite{houri2019limit}.\\
\indent\indent\ To use the second fluxural mode as a modal pump, a self-sustained oscillator is implemented by constructing a gain feedback loop around mode 2. The LDV output is passed through a filter-amplifier (NF3627), and then fed back into the device as shown schematically in Fig.~2(a). A band-pass filter with a 48 dB/Octave slope is selected, the center frequency of the filter and the amplifier gain are swept electronically during measurements. The band-pass filter is inserted to insure that only the second mode goes into self-oscillation, and all other modes are either driven or parametrically pumped via modal interactions with the second mode.\\
\indent\indent\ Upon increasing the gain of the feedback loop, the observed quality factor of the second mode increases until reaching self-oscillation for an effective loop gain of $\sim$ 25, as shown in Fig.~2(b). Conversely, the line width of the second mode decreases with increasing loop gain at a rate of $\sim$10 Hz/unit g, where g is the effective loop gain given as $g(\omega) = G(\omega).cos(\theta(\omega))$ where $G(\omega)$ and $\theta(\omega)$ are the nominal amplifier gain and the filter-induced feedback phase lag, respectively. Note that because of the band-pass filter both $G$ and $\theta$ are frequency dependent (see supplementary material).\\
\indent\indent\ Once the second mode enters the regime of self-oscillation, its frequency can be tuned without the need for external forcing, unlike what is usually the case for oscillators \cite{houri2017direct}. This is performed by keeping the gain of the feedback loop constant and slowly changing the center frequency of the band-pass filter, the frequency of the self-oscillating mode is thus pulled along with that of the band-pass filter, as shown in Fig.~2(c) bottom panel.\\
\indent\indent The oscillator exhibits a wide tuning range ($\approx$ 30\%), and although a Duffing-van der Pol oscillator can have its frequency tuned (see supplementary material), that requires changing the gain of the system. The observed frequency shift takes place even for a constant nominal gain, as can be seen by the yellow data points in the lower panel of Fig.~2(c). To understand this behaviour modal interactions need to be considered. However, since accouting for all five modes renders the system intractable, and since at the onset of self-oscillations the frequency is near the resonance frequency, it is reasonable to limit the analysis to the second and fourth modes (the nearest internally resonant mode), and observe how the experimental response deviates from this simplified model.\\
\indent\indent By considering only the two modes, linearizing the higher frequency one and applying the rotating frame approximation where the modal displacement takes the form $\mathrm{x_j(t)=A_j(t)cos(\frac{\textit{j}}{2}\times\omega t+\phi_j(t))}$, where $\mathrm{A_j(t)}$ and $\mathrm{\phi_j(t)}$ are slowly varying amplitude and phase envelopes of the \textit{j}th mode \cite{greywall1994evading}, and $\omega$ is the second mode oscillation frequency. Introducing these into Eq.(1) (after dropping the forcing term) gives an explicit form for the self-oscillating second mode, expressed as:\\
\begin{multline}
(\frac{3\alpha_2}{8}E_2 + \delta_2)^2 + (\frac{\omega\beta_2}{8}E_2 + \frac{\omega\gamma_2}{2})^2 =   9(\frac{g_{24}}{8})^4\frac{E_2^4}{\Delta}
\label{eqn:Eq2}
\end{multline}
\indent\indent\ where $E_2 = A_2^2$, $\delta_2 = \omega_{0,2}^2-\omega^2$, $\Delta = (\omega_{0,4}^2 - 9\omega^2)^2 + (\frac{3\omega\gamma_4}{2})^2$, and $\gamma_2 <0$. Note that Eq.~(2) starts as a quintic polynomial for which $A_2=0$ is always a solution, but not necessarily a stable one.\\
\begin{figure}[t]
	\graphicspath{{Figures/}}
	\includegraphics[width=85mm]{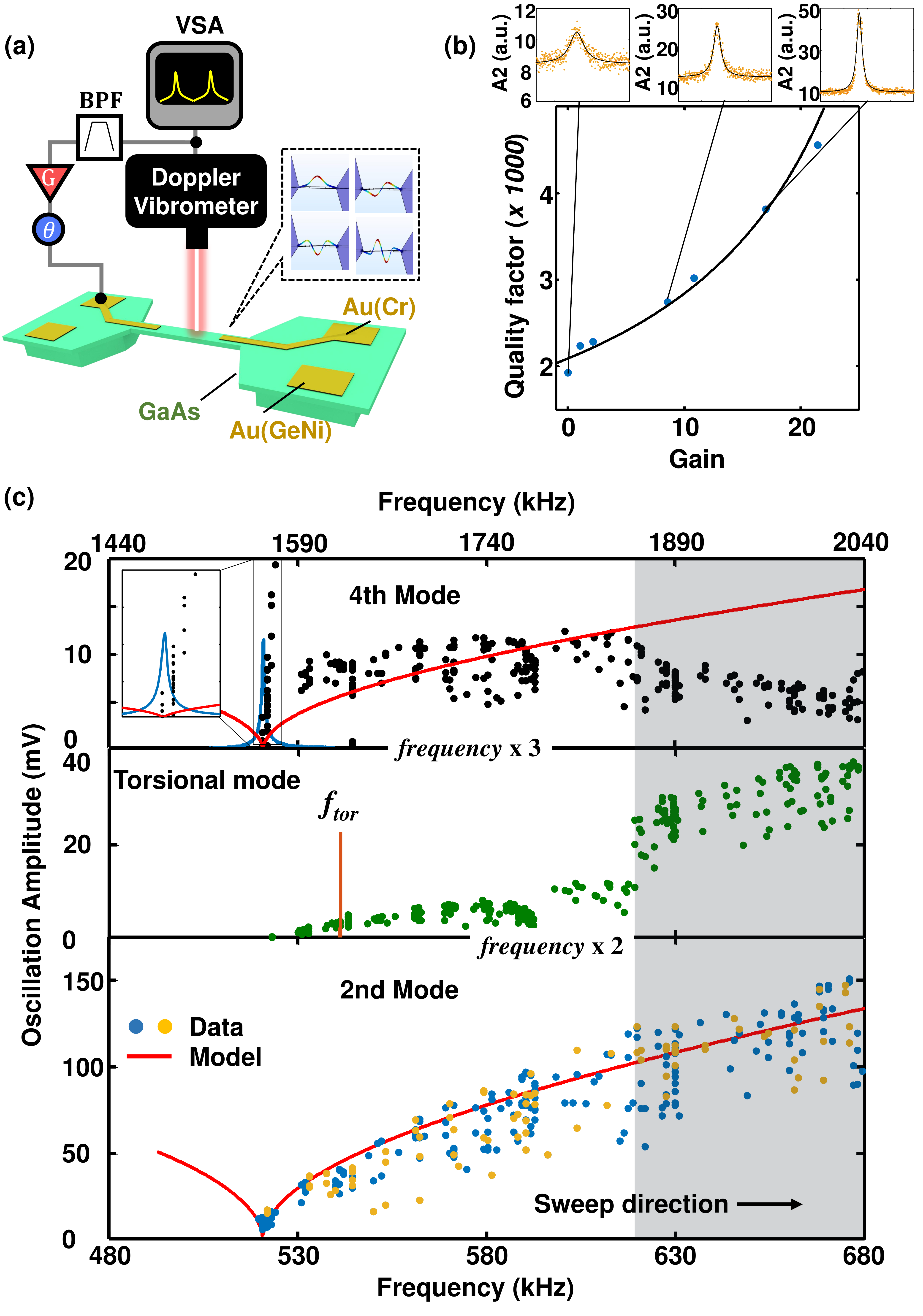}
	\caption{Schematic representation of the gain feedback loop used to induce self-oscillations in the second mode (a). (b) Plot of the quality factor as a function of nominal loop gain, indicating a self-oscillation threshold of $\sim$ 25. The insets show the change in the Brownian response as a function of gain. (c) Amplitude-frequency response of the device showing the post self-oscillations response of the second flexural (lower panel), torsional (mid-panel), and fourth flexural (top panel) modes. The  scaling behavior as expected from the model is shown for comparison (red trace bottom and top panels). All the yellow data points (bottom panel) are obtained for a constant loop-gain of 250, blue data points are for various values of loop gain. The resonance frequency of the torsional mode (red line, middle panel) and the response of the fourth mode (blue trace, top panel) are shown for visual comparison. Both the torsional and the fourth flexural modes change their scaling behavior at the onset of the 2:1 parametric interactions between the second and the first flexural modes (shaded area).}
\end{figure}
\indent\indent\ Using the fitted parameters and the simplified model, the amplitude-frequency response is simulated and shown in Fig.~2(c). The qualitative scaling behavior shows good agreement with experimental results, knowing that the quantitative response is left as a free parameter. The model predicts a negative detuning branch that was not observed experimentally, along with a relatively large scatter in experimental data that can not be solely attributed to experimental factors.\\
\indent\indent\ Further insight is obtained by observing the response of the higher modes, also shown in Fig.~2(c). The fourth mode shows a significant departure from the model near its natural frequency, where small errors are magnified by resonance. However as the oscillation frequency is pulled, the fourth mode approaches the behavior described by the model, thus despite its simplicity, the model gives an approximate qualitative description of the system.\\
\indent\indent\ However, the response changes drastically around $\sim620$ kHz, and the torsional mode that is not included in the model simultaneously shows a large change in its response. Since at this frequency the self-oscillating mode is in a near 2:1 frequency ratio with the first flexural mode, parametric interactions between the two take place. This can be seen in Fig.~3(a), where as a consequence of a slight frequency pulling of the oscillator, the first mode enters a regime of parametric oscillations.\\
\indent\indent\ The onset of the 2:1 parametric interactions, and the qualitative change in the behavior of the torsional and the fourth modes indicate a global change in the system dynamics. This implies that the simplification has reached its limits, and beyond this point even qualitative understanding of the system dynamics requires multimode analysis.\\
\indent\indent\ It is possible however to fit a value for the three-wave mixing mode coupling component between the first and second modes, by treating the second mode as a parametric pump and observing the threshold for parametric resonance as a function of amplitude and detuning, as shown in Fig.~3(b). Since below the threshold Duffing nonlinearity and internal resonance can be neglected, the only nonlinear term of consequence is the mode coupling-induced parametric interaction term, thus Eq.~(1) can be modified to the following form:\\
\begin{multline}
\mathrm{\ddot{x}_1 + \gamma_1\dot{x}_1 + (\omega_{1,0}^2+\kappa_{12}A_2cos(\omega t+\phi_2))x_1 = 0}
\label{eqn:Eq3}
\end{multline}
\indent\indent Equation 3 is similar to a canonical Mathieu equation \cite{nayfeh2008nonlinear}. Using Eq.~(3) and fitting the experimental threshold data for the parametric resonance, a value for $\kappa_{12}$ is obtained, which in turn is used to estimate the area of parametric oscillations. These are shown in Fig.~3(b) along with the actual measured parametric oscillation area. While the Mathieu equation fits the onset of oscillations, it fails to predict its limits. Again, an indication that a more global treatment of the system is necessary once all 5 modes are interacting.\\
\indent\indent\ Further plots of different modal amplitudes (not shown) show large scatter in their distribution, but do not provide further understanding. The phase on the other hand offers hints regarding the behaviour of the different modal components. By plotting the modal amplitudes versus that of the second mode, one would expect Lissajous-type figures since the frequency of the various modes are in a n:2 ratio with that of the second mode, an example of these is shown in Fig.~3(c). And whereas the torsional and the fourth modes show a stable phase relation, most surprisingly, the first mode, which is undergoing parametric oscillations, does not. This is counter intuitive since a constant $\pi$ phase difference can be expected for a parametric oscillator.\\
\indent\indent\ Extracting the phase difference of the modal components for the entire data set, and plotting these as a function of the second mode oscillation frequency reveals more of the global system dynamics, as shown in Fig.~3(d). The system clearly undergoes a transition at the onset of the 2:1 parametric oscillations. Before the oscillations, modes 1 and 3 have zero amplitudes (as expected) and the fourth and the torsional modes are excited via four-wave mixing and three-wave mixing, respectively. Interestingly, these two modes do not show a very constant phase relation, i.e. the phase seems not to show a clear trend with the frequency, in addition the standard deviation of the phase difference plotted as error bars  shows in some, but not all, of the cases a large component, equally indicative of a time-dependent phase difference.\\
\indent\indent\ Upon the onset of parametric oscillations, the fourth and the torsional modes have their phases anchored, they show indications of a tristable phase with a slight frequency-dependence and negligible deviations, corresponding to what is already observed in Fig.~3(c). Modes 1 and 3, on the other hand, confirm the lack of any coherent phase difference as their mean value fluctuates, and the associated error bars span the entire $[-\pi\quad\pi]$ range. Equally interesting to note, is that if the first and third modes oscillations are plotted against each other, no constant phase difference is observed. This suggests that the third mode is not a high frequency copy of the first mode, but more complicated interactions with the even modes are at play.\\
\indent\indent\ The most reasonnable hypothesis for the lack of coherence is the presence of a time-dependent phase component, most likely caused by either a Hopf or a SNIPER bifurcation \cite{houri2019limit,czaplewski2018bifurcation}. However, such bifurcations usually result in the appearance of side bands around the main modal oscillation frequency, and such side bands were conspicuously missing from the data set.\\
\indent\indent\ Regardless of the underlying mechanism responsible for the phase behavior, the experimental means presented here provide a novel and convenient method for the formation of multimodal frequency combs. With the additional prospect of having the even and odd modes in a constant frequency ratio, but with an unlocked phase.\\
\begin{figure}[t]
	\graphicspath{{Figures/}}
	\includegraphics[width=85mm]{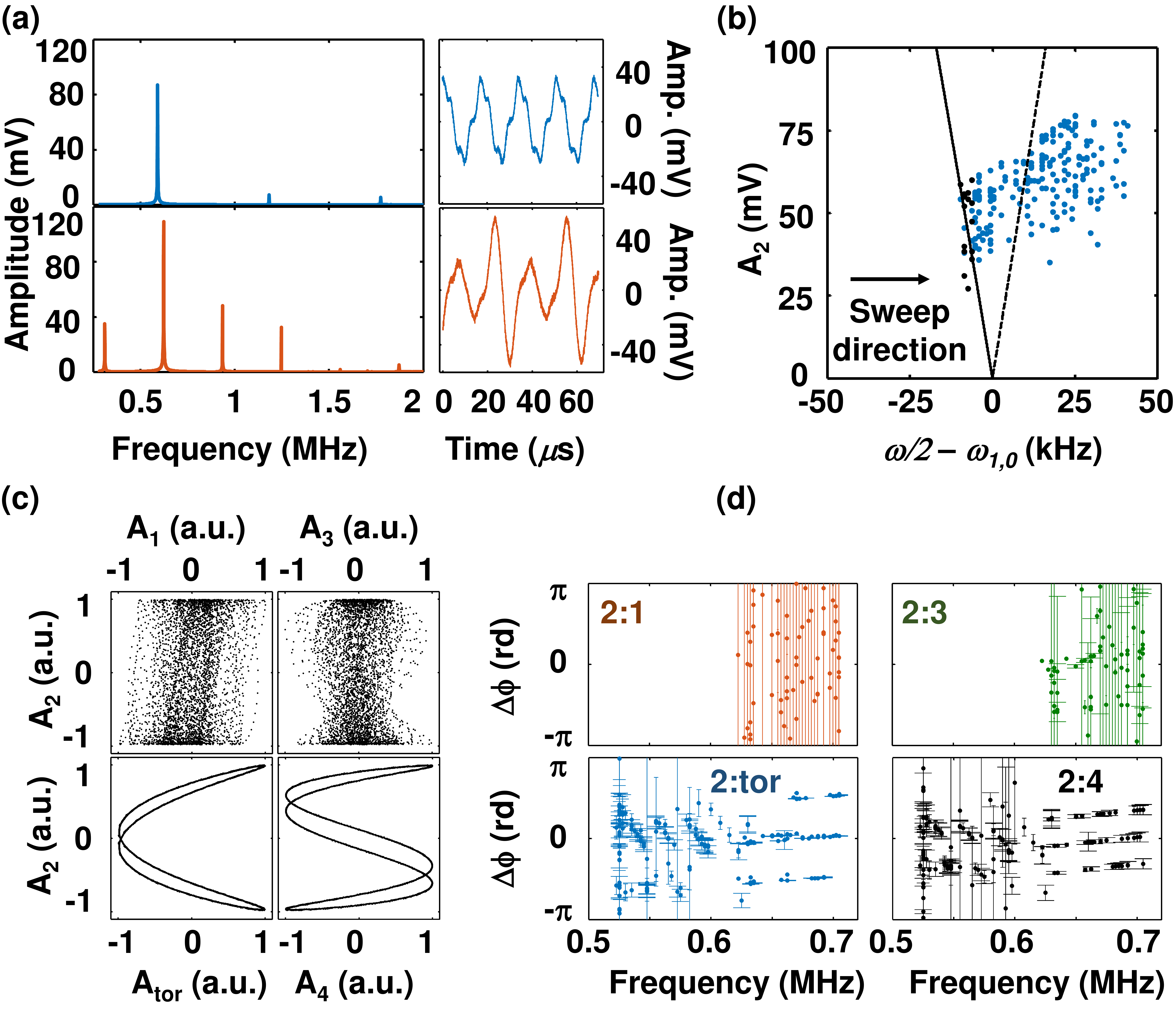}
	\caption{Frequency domain plot of the oscillator before (blue) and after (red) the onset of 2:1 parametric oscillations (a), along with a time domain sample showing the change in the waveform output before (blue) and after (red) parametric resonance (right panels). (b) Experimental data showing the loci of the parametric resonance points as a function of the second mode amplitude and detuning (blue and black points). The data points corresponding to the onset of parametric resonance are shown in black, the parametric resonance area for a typical parametric oscillator is delineated with the black lines for the forward (solid line) and the backward frequency sweeps (dashed line), respectively. (c) Lissajous figures obtained by plotting second mode amplitude against the first mode (top left), third mode (top right), fourth mode (bottom right), and the torsional mode (bottom left). The first two plots indicate the presence of an additional time-dependent phase difference between the even and odd modes (data set shown correspond to the red trace in (a)). (d) Plots as a function of the oscillator frequency of the phase difference between mode 2 and the first mode (red), third mode (green), fourth mode (black), and the torsional mode (blue). The first and third modes show no constant phase and a large deviation corresponding to a time-dependent phase difference. The fourth and torsional modes show a phase tristability in the region of parametric oscillations.}
\end{figure}
 \bibliographystyle{apsrev}
\bibliography{aipsamp}
\end{document}